\documentclass[aps,prd, twocolumn,superscriptaddress,nofootinbib,preprintnumbers]{revtex4-1}
\usepackage{mathrsfs}
\usepackage{indentfirst}
\usepackage{amsmath}
\usepackage{bigstrut}
\usepackage{amssymb}
\usepackage{array}
\usepackage{multirow}
\usepackage{etoolbox}
\usepackage{graphicx}
\usepackage{amsthm}
\usepackage{slashed}
\usepackage{array}
\usepackage{epstopdf}
\usepackage{graphicx}
\usepackage{fancyhdr}
\usepackage[table,xcdraw]{xcolor}
\usepackage{float}
\usepackage{subfigure}
\usepackage{url}
\usepackage{hyperref}
\graphicspath{{Figure/}}

\begin{document}
\title{$W$-Boson Mass, Electroweak Precision Tests and SMEFT}

\author{JiJi Fan}
\email{jiji\_fan@brown.edu}
\affiliation{Department of Physics, Brown University, Providence, RI, 02912, USA}
\affiliation{Brown Theoretical Physics Center, Brown University, Providence, RI, 02912, USA}

\author{Lingfeng Li}
\email{lingfeng\_li@brown.edu}
\affiliation{Department of Physics, Brown University, Providence, RI, 02912, USA}

\author{Tao Liu}
\email{taoliu@ust.hk}
\affiliation{Department of Physics, The Hong Kong University of Science and Technology, Clear Water Bay, Kowloon, Hong Kong S.A.R., PRC}

\author{Kun-Feng Lyu}
\email{lyu00145@umn.edu}
\affiliation{School of Physics and Astronomy, University of Minnesota, Minneapolis, MN 55455, USA}

\unitlength = 1mm

\begin{abstract}

Recently the CDF collaboration at the Tevatron reported a significant discrepancy between the direct measurement of the $W$-boson mass and its Standard Model (SM) prediction based on electroweak precision tests (EWPTs). In this paper we explore the potential origin of this discrepancy from physics beyond the SM. Explicitly, we work on a set of six-dimensional operators in the SM effective field theory (SMEFT) which are relevant to the EWPTs. By fitting to the data, we demonstrate that an upward shift in $m_W$ is driven by the operator $\mathcal{O}_{T}=\frac{1}{2}(H^{\dagger}\overset{\text{$\leftrightarrow$}}{D}_{\mu}H)^2$ with a coefficient $c_T ({\rm TeV}/\Lambda)^2 \gtrsim 0.01$. This suggests that the new physics scale favored by the CDF data should be multiple TeV for tree-level effects and sub TeV for loop-level effects. One simple example is to introduce a hypercharge-free electroweak triplet scalar which can raise the $c_T$ value at tree level. We also study the potential to further test the relevant SMEFT by measuring Higgs-coupling, $m_W$ and other EWPTs at future circular $e^-e^+$ colliders.

\end{abstract}

\maketitle

\section{Introduction}

The discovery of the Higgs boson~\cite{Aad:2012tfa,Chatrchyan:2012xdj} marks a great success of the Standard Model (SM) in particle physics. It opens the door to explore mass origin of the  massive elementary particles and the underlying physics for electroweak symmetry breaking (EWSB), and thus has far-reaching impacts for the development of particle physics. The measurements of the $W$ boson can also play a significant role in this direction. As a direct consequence of EWSB, the $W$-boson mass ($m_W$) could be determined by the $Z$-boson mass ($m_Z$) and the Weinberg angle ($\theta_W$). Thanks to the high precision achieved in electroweak precision tests (EWPTs), $m_W$ has been well-constrained in the SM. We denote this measured value as $m_W^{\rm EW}$. Then any deviation from $m_W^{\rm EW}$ in the direct measurements of $m_W$ may serve as a signal of physics beyond the SM (BSM). 

Recently, the Collider-Detector-at-Fermilab (CDF) collaboration at Tevatron reported its updated measurement of $m_W$~\cite{doi:10.1126/science.abk1781}. With $\sim 4.24\times 10^6$ semileptonic $W$-decay events, this measurement yields $m_W=80.4335 \pm 0.0094$~GeV~\cite{doi:10.1126/science.abk1781}. Here a combination of the almost quadrupled statistics and the improvements in systematic effects such as parton distribution function and lepton resolution leads to a greatly reduced uncertainty of $\sim9.4$~MeV. This precision is approximately halved compared to that of the previous CDF measurement~\cite{CDF:2012gpf} and also exceeds the current experimental average ($\sim 12$~MeV)~\cite{ParticleDataGroup:2020ssz}. Such a $W$-boson mass results in a tension of $\sim 7\sigma$ with the latest EW global fit of $m_W^{\rm EW}=80.3545 \pm 0.0059$~GeV~\cite{deBlas:2021wap}. Caution should be taken when we attend to make an interpretation of this tension. There is a significant tension between the new CDF result and the direct measurements performed by the D0 collaboration at the Tevatron~\cite{D0:2012kms} and the ATLAS/LHCb collaborations at the Large Hadron Collider (LHC)~\cite{ATLAS:2017rzl,LHCb:2021bjt}, while the latter ones are in good agreement with $m_W^{\rm EW}$. This indicates that a better understanding of the uncertainties in these measurements is needed, to ensure a solid interpretation of the data. Notably, dedicated (HL-)LHC runs with a low instantaneous luminosity and upgraded detectors may also measure $m_W$ with a precision $\lesssim 10$~MeV~\cite{ATLAS:2018qzr,Azzi:2019yne}, falsifying or strengthening the new CDF result.

With these subtleties in mind, we would explore the potential origin of this discrepancy from the BSM physics. For this purpose, we will simply assume that the direct measurements of $m_W$ will later converge to a value close to the CDF one~\cite{doi:10.1126/science.abk1781}. It is already known that the BSM scenarios exist where the $m_W$ value can deviate from the SM prediction without heavily disturbing other electroweak precision observables (EWPOs)~\cite{Berthier:2015oma,Bjorn:2016zlr,Corbett:2021eux}. The examples include the singlet~\cite{Lopez-Val:2014jva} or doublet~\cite{Lopez-Val:2012uou} extensions of the SM Higgs sector where the corrections to $m_W$ enter at one-loop level. A significant correction to $m_W$ can be also achieved in supersymmetry with the sfermion-loop contributions~\cite{Heinemeyer:2013dia,Diessner:2019ebm,Bagnaschi:2022qhb}. Moreover, a positive correction to $m_W$ is possible via a tree-level mixing between the SM $Z$ boson and the BSM ``dark photon"~\cite{Curtin:2014cca,Alguero:2022est}. In composite Higgs theories, $m_W$ can be shifted away from its SM prediction due to the UV-scale suppressed interactions~\cite{Bellazzini:2014yua}.   

In this paper, we will try to explore this discrepancy in the framework of the SM effective field theory (SMEFT) and make a prediction on its impacts for the ongoing and upcoming measurements at colliders. In the SMEFT, we parametrize the leading effects of the BSM physics above the EW scale with a set of dimension-six (6D) operators
\begin{equation}
\mathcal{L}_{\text{eff}} = \mathcal{L}_{\text{SM}} + \sum_{i} \frac{c_i}{\Lambda^2} \mathcal{O}_i  \ ,
\end{equation}
where $\mathcal{L}_{\text{SM}}$ defines the SM, and $c_i$ and $\Lambda$ denote dimensionless Wilson coefficients and the cutoff scale of the BSM physics, respectively. Instead of presenting a comprehensive study on the whole set of 6D operators, we will focus on the key ones which are closely related to the $W$-boson physics. We would view this as a first step for more refined analyses later.

We organize this paper in the following way. We will introduce the analysis formalism in Section~\ref{sec:formalism}, which was originally developed in~\cite{Chiu:2017yrx}. The analysis results of the SMEFT and the predictions for the ongoing and future tests will be presented in Section~\ref{sec:results}. We will conclude in Section~\ref{sec:conclusions}.

\section{Analysis Formalism}
\label{sec:formalism}

The observables to be considered in this paper are summarized in Tab.~\ref{tab:data-lep}, which include $m_W$, the EWPOs and one Higgs observable. Below is the list of the EWPOs:

\begin{equation}
\begin{split}
 &R_b = \dfrac{\Gamma_b}{\Gamma_{\text{had}}}, \quad R_{\ell}=\dfrac{\Gamma_{\text{had}}}{\Gamma_{\ell}},  \quad   A_f =  \dfrac{2 g_V^f}{g_V^f + g_A^f}, \\
& \sin^2 \theta_{\text{eff}}^{\text{lep}}= \dfrac{1}{4} \left( 1- \dfrac{g^l_V}{g^l_A} \right), \quad \Gamma_Z, \quad \sigma_{\rm had}^0 = \frac{12 \pi}{m_Z^2}\frac{\Gamma_e \Gamma_{\rm had}}{\Gamma_Z^2},\\
& A_{FB}^f = \dfrac{3}{4} A_e A_f  \quad (f = b, \ell),\quad \Gamma_W, \quad \text{BR}_{W\to {\rm had}}  \ .
\end{split} 
\end{equation}
Here $\Gamma_f = \Gamma(Z \rightarrow f \bar{f})$, $\Gamma_{\rm had} = \Gamma(Z \to \text{hadrons})$ and $\Gamma_Z$ are the partial and total decay widths of the $Z$ boson.  $g_{V(A)}^f$ is the (axial-)vector couplings of fermion $f$. $\Gamma_W$ and $\text{BR}_{W\to {\rm had}}$ denote the total decay width and branching fraction of hadronic decays of $W$ boson.  Note that we take the experimental average at the LHC and Tevatron as the value of $\sin^2 \theta_{\text{eff}}^{\text{lep}}$, which is thus uncorrelated with the asymmetry observables ($A$'s) measured at LEP.

\begin{table*}
  \centering
\scalebox{0.76}{
    \begin{tabular}{c|c|c|c|c}
    \hline
    \hline
    && Experimental Measurement &  Precisions at Future $e^-e^+$ &Theoretical Prediction  \\
    \hline
    \hline
 \multirow{3}{*}{EW Inputs}  &  $m_Z$(GeV) & $91.1875 \pm 0.0021$\cite{ALEPH:2005ab} & $\pm1.0 \times 10^{-4}$\cite{FCC:2018byv} & / \\
    \cline{2-5}
  &  $G_F(10^{-10} \text{GeV}^{-2})$    & $1166378.7 \pm 0.6$ \cite{Patrignani:2016xqp} & $\pm0.6$ & / \\
    \cline{2-5}
    \cline{2-5}
  &  $ \alpha^{-1}(m_Z)$ & $127.952 \pm 0.009$ ~\cite{ParticleDataGroup:2020ssz}&  $\pm0.003$~\cite{FCC:2018byv} & / \\
    \cline{2-5}
    \hline
    \hline
\multirow{14}{*}{ Observables}  &  $m_W$(GeV)   & 80.4335 $\pm$ 0.0094 \cite{CDF:2022hxs} &  $\pm 3.9 \times 10^{-4}$ \cite{FCC:2018byv} &  80.3545 $\pm$ 0.0059 ($\pm$ 0.0014)\\
    \cline{2-5}
  &  $A_b$    & 0.923 $\pm$ 0.020 \cite{ParticleDataGroup:2020ssz,ALEPH:2005ab}  & $\pm2.9 \times 10^{-3}$ \cite{FCC:2018byv} &  0.934727 ($\pm 0.000014$)\\
    \cline{2-5}
  &  $A_\ell$ ($P_{\tau}$)    & 0.1465  $\pm$ 0.0033 \cite{ParticleDataGroup:2020ssz,ALEPH:2005ab}  & $\pm4.3 \times 10^{-5}$ \cite{FCC:2018byv}  & 0.14692 ($\pm 0.00018$)\\
    \cline{2-5}
  &  $A_\ell$ (SLD)    & 0.1513  $\pm$ 0.0021 \cite{ParticleDataGroup:2020ssz,ALEPH:2005ab}  & / & 0.14692 ($\pm 0.00018$)\\
    \cline{2-5}
   & $R_b$    & 0.21629 $\pm$ 0.00066 \cite{ParticleDataGroup:2020ssz,ALEPH:2005ab} & $\pm6.0\times 10^{-5}$  \cite{FCC:2018byv}& 0.21588 ($\pm 0.00005$) \\
    \cline{2-5}
  &  $R_\ell$   & 20.767 $\pm$ 0.025 \cite{ParticleDataGroup:2020ssz,ALEPH:2005ab} & $\pm1.0 \times 10^{-3}$ \cite{FCC:2018byv}  & 20.7464 ($\pm 0.0020$)\\
    \cline{2-5}
   &  $A_{\rm FB}^b$   & 0.0996  $\pm$ 0.0016 \cite{ParticleDataGroup:2020ssz,ALEPH:2005ab} & $\pm3.0 \times 10^{-4}$ \cite{FCC:2018byv}  & 0.10300 ($\pm$ 0.00013) \\
    \cline{2-5}
       &  $A_{\rm FB}^\ell$   & 0.0171  $\pm$ 0.0010 \cite{ParticleDataGroup:2020ssz,ALEPH:2005ab} & $\pm2.0 \times 10^{-4}$ \cite{FCC:2018byv}  & 0.01619 ($\pm 0.00040$)\\
    \cline{2-5}
  &  $\Gamma_Z$(GeV) & 2.4955 $\pm$  $0.0023$~\cite{ParticleDataGroup:2020ssz} &  $ \pm2.5 \times 10^{-5}$ \cite{FCC:2018byv} & 2.49414 ($\pm 0.00019$)\\
      \cline{2-5}
  &  $\sigma_{\rm had}^0$(nb) & 41.480 $\pm$  $0.033$~\cite{Janot:2019oyi} & $\pm4.0 \times 10^{-3}$ \cite{FCC:2018byv}  &41.4929 ($\pm 0.0032$)\\
      \cline{2-5}
  &  $\Gamma_W$(GeV) & 2.085 $\pm$  $0.042$~\cite{ParticleDataGroup:2020ssz}
  &  $\pm1.2 \times 10^{-3}$ \cite{FCC:2018byv} & 2.08782 ($\pm 0.00011 $)\\ \cline{2-5}
 &  BR$_{W\to {\rm had}}$ & 0.6741 $\pm$  $0.0021$~\cite{ParticleDataGroup:2020ssz}
   &  $\pm 6.4 \times 10^{-5}$ \cite{dEnterria:2016rbf} & 0.6748 ($\pm 0.00010$)\\ \cline{2-5}
   & $\sin^2 \theta_{\text{eff}}^{\text{lep}} (10^{-5})$  & $(23143 \pm 25)$~\cite{ATLAS:2015ihy,LHCb:2015jyu,CMS:2018ktx,ATLAS:2018gqq,CDF:2018cnj}  &  $\pm0.31$ \cite{FCC:2018byv} & 23153.4  ($\pm 2.3$) \\\cline{2-5}
   & $\mu_{ggh}^{\gamma\gamma}/\mu_{\rm SM}$ & 1.02 $\pm$ 0.11 \cite{Rossi:2022alp} & / & 1
   \\\cline{2-5}
   & BR($h \to \gamma\gamma$)/BR($h \to \gamma\gamma$)$_{\rm SM}$ & / & $\pm0.0032$ \cite{FCC:2018byv}& 1\\
    \hline
    \hline
    \end{tabular}}%
  \caption{Input parameters, EWPOs and Higgs observable for the analyses with current data and the sensitivity projections at a future circular $e^-e^+$ collider. For the Higgs observable, we use $\mu_{ggh}^{\gamma\gamma}/\mu_{\rm SM}$ in the former case and BR($h \to \gamma\gamma$)/BR($h \to \gamma\gamma$)$_{\rm SM}$ in the latter case. These two are equivalent in our framework. For the uncertainties of theoretical predictions, we consider the one of $m_W$ only in the former case, but the ones for all except the Higgs observable in the latter case. Their values have been listed outside and inside the brackets in the last column, respectively. }
  \label{tab:data-lep}%
\end{table*}%

Totally six CP-even 6D operators will be considered in this study, which include  
\begin{eqnarray}
 \label{eq:opes}
\mathcal{O}_{WB}&=&gg'H^{\dagger}{\sigma}^aHW^{a}_{\mu \nu}B^{\mu \nu}  , \nonumber \\   \mathcal{O}_{T}&=&\frac{1}{2}(H^{\dagger}\overset{\text{$\leftrightarrow$}}{D}_{\mu}H)^2 \, , \nonumber \\
\mathcal{O}^{(3)l}_{L}&=&(iH^{\dagger}{\sigma}^{a}\overset{\text{$\leftrightarrow$}}{D}_{\mu}H)(\bar{L}_L{\gamma}^{\mu}{\sigma}^{a}L_L), \nonumber\\
\mathcal{O}^{(3)l}_{LL}&=&(\bar{L}_L{\gamma}_{\mu}{\sigma}^{a}L_L)(\bar{L}_L{\gamma}^{\mu}{\sigma}^{a}L_L), \nonumber\\
\mathcal{O}^l_L&=&(iH^{\dagger}\overset{\text{$\leftrightarrow$}}{D}_{\mu}H)(\bar{L}_{L}{\gamma}^{\mu}L_L) ,  \nonumber \\  \mathcal{O}^e_R&=&(iH^{\dagger} \overset{\text{$\leftrightarrow$}}{D}_{\mu}H)(\bar{l}_{R}{\gamma}^{\mu}l_R) \, . \nonumber   
\end{eqnarray}
Here we work in the Warsaw basis~\cite{Grzadkowski:2010es}.\footnote{Our $\mathcal{O}_{T}$ is slightly different from $\mathcal{O}_{HD} = (H^\dagger D_\mu H)^* (H^\dagger D_\mu H)$ in the original Warsaw basis. They are related as $\mathcal{O}_T=-2\mathcal{O}_{HD}-\frac{1}{2}\partial_\mu\left(H^\dagger H \right) \partial^\mu\left(H^\dagger H \right)$. } These operators can affect the EW physics by renormalizing wave functions, shifting the definition of EW parameters and the SM couplings, and introducing new vertices. We will follow the linear formalism developed in~\cite{Chiu:2017yrx} to analyze these effects. Strictly speaking, there are additional CP-even 6D operators that could affect EWPTs such as Higgs coupling to quarks~\cite{Ellis:2020unq}. Yet our choice already suffices for the leading-order discussions on the potential BSM physics to explain the new CDF result, and explore other possible measurable consequences.

To derive the relative shifts caused by these operators in the observables listed Eq.~(\ref{eq:opes}), we need three basic input parameters: the $Z$-boson mass $m_Z$, the Fermi constant $G_F$, and the fine-structure constant $ \alpha^{-1}(m_Z)$. Their experimental values and uncertainties are presented in Tab.~\ref{tab:data-lep}. For the other input parameters that contribute to the EW predictions in the SM at one or higher-loop levels, namely $m_h$, $m_t$, $\alpha_s(m_Z)$, their values are taken to be the same as those in~\cite{deBlas:2021wap}. Following the procedures described in~\cite{Fan:2014vta,dEnterria:2016rbf,Dubovyk:2018rlg} with the updated parameter values, we calculate the SM values for the EWPOs which agree with~\cite{deBlas:2021wap} decently, with a difference $\lesssim 1\sigma$. As for theoretical uncertainties, only the one for $m_W$ is relevant. The overall uncertainty of $\sim 5.9$~MeV for the $m_W$ prediction is evenly contributed by the uncertainty of $m_t$ and the high-order corrections. By taking this effect into consideration, the predicted uncertainties for these observables also match well with~\cite{deBlas:2021wap}.

To show how the 6D operators affect these EWPOs, let us consider $m_W$ as an example. The $W$-boson mass receives the contributions from these operators via the EW-parameter shifts only, which yields  
\begin{eqnarray}
\dfrac{\Delta m_W}{m_W} = \dfrac{\delta g_Z}{g_Z} -\dfrac{\sin \theta_W}{\cos \theta_W} \delta \theta_W -\dfrac{1}{2} \dfrac{\delta G_F}{G_F} \ .
\end{eqnarray}
 Here $\theta_W$ and $g_Z$ are defined as 
  \begin{eqnarray}
 \sin 2 \theta_W &=&  \left( \dfrac{4 \pi \alpha }{\sqrt{2} {G}_{F} m^2_{Z}} \right)^{1/2}, \nonumber \\  g_Z &=& \dfrac{g}{\cos \theta_W} = \dfrac{ 4 \sqrt{\pi \alpha}}{\sin 2 \theta_W} = 2 (\sqrt{2} G_F m_Z^2)^{1/2} \ , 
\end{eqnarray}
where $g$ is the SM $SU(2)$ coupling. Then we have 
\begin{eqnarray}
 \delta \theta_W &= &\dfrac{\sin \theta_W \cos \theta_W}{2(\cos^2 \theta_W-\sin^2 \theta_W)} \left(\dfrac{\delta \alpha}{\alpha} - \dfrac{\delta G_F}{G_F} - \dfrac{2\delta m_Z}{m_Z} \right) ,\nonumber \\
   \dfrac{\delta g_Z}{g_Z} &=& \dfrac{1}{2}\dfrac{\delta G_F}{G_F} + \dfrac{\delta m_Z}{m_Z}  \ .
\end{eqnarray}
The parameter shifts depend on the Wilson coefficients of the 6D operators as 
\begin{eqnarray}
\dfrac{\delta m_Z}{m_Z} &=& - \delta Z_Z + \dfrac{c_T v^2}{2\Lambda^2},  \quad
\dfrac{\delta G_F}{G_F} = \dfrac{2 (c^{(3)l}_{LL} - c_L^{(3)l}) v^2}{\Lambda^2}, \nonumber\\
\dfrac{\delta \alpha}{\alpha} &=& -2 \delta Z_A  \ ,
\end{eqnarray}
where $v = 246$GeV is the SM Higgs VEV. 
$\delta Z_Z$ and $\delta Z_A$ are field renormalization factors induced by $\mathcal O_{WB}$. They are given by 
\begin{equation}
\begin{split}
 \delta Z_Z & = \dfrac{v^2}{\Lambda^2} \cos \theta_W \sin \theta_W g g' c_{WB},  \\
 \delta Z_A &= - \dfrac{v^2}{\Lambda^2} \cos \theta_W \sin \theta_W g g' c_{WB} \ ,
\end{split}
\end{equation}
where $g^\prime$ is the SM $U(1)_Y$ coupling. Combining these derivations, one will find the dependence of $m_W$ on the Wilson coefficients of these 6D operators at linear level.  
Generalizing this discussion, we have 
\begin{eqnarray}
\frac{\Delta \mathcal O}{\mathcal O} &=& C_{\mathcal O}^{WB} \dfrac{c_{WB}}{\Lambda^2} + C_{\mathcal O}^{T} \dfrac{c_{T}}{\Lambda^2}   + C_{\mathcal O}^{3L}\dfrac{c_{L}^{(3)l}}{\Lambda^2} \nonumber \\
&+&C_{\mathcal O}^{3LL} \dfrac{c^{(3)l}_{LL}}{\Lambda^2}  
+C_{\mathcal O}^{L} \dfrac{c^{l}_{L}}{\Lambda^2}  +C_{\mathcal O}^{R} \dfrac{c^e_R}{\Lambda^2}     \label{eq:ge}
\end{eqnarray}
for all EWPOs. The coefficients $C_{\mathcal O}$'s are summarized in Tab.~\ref{tab:coeff}.

\begin{table}[ht]
  \centering
\scalebox{0.9}{
    \begin{tabular}{c|c|c|c|c|c|c}
    \hline\hline
$\mathcal {O}$ & $C_{\mathcal O}^{WB}$ &  $C_{\mathcal O}^T$  & $C_{\mathcal O}^{3L}$ & $C_{\mathcal O}^{3LL}$ & $C_{\mathcal O}^{L}$ & $C_{\mathcal O}^{R}$ \\ 
   \hline
$m_W$ &  -0.0111&0.0433&-0.0264&0.0264&/& /\\ 
   \hline
$A_b$ &-0.00781&0.0142& -0.0285 & 0.0285 &/ & / \\ 
   \hline
  $R_b$ &0.00189&-0.00345& 0.00691 &-0.00691 & / & /\\ 
   \hline  
 $R_\ell$ &-0.00969 &0.0177& -0.159& 0.0353& -0.124& 0.109 \\ 
   \hline 
   $A_{FB}^{\ell}$ &  -1.01 & 1.84 &  -2.41 & 3.69  & 1.28 & 1.46 \\ 
    \hline
   $A_\ell$ & -0.583& 1.06 & -1.38 &  2.13  & 0.739 & 0.843 \\ 
   \hline  
$A_{FB}^b$ & -0.625 & 1.14 & -1.50 & 2.28 & 0.784 & 0.894 \\ 
   \hline
   $\Gamma_Z$ &-0.0112& 0.079&  -0.121 & 0.158 & -0.0113& -0.0113\\     \hline
   $\sigma_{\rm had}^0 $ & 0.00142 & -0.00259 & 0.0572 & -0.00519 & 0.152 & -0.0895 \\     \hline
   $\Gamma_W$ &-0.0322& 0.126&  -0.174 & 0.193 & /& /\\     \hline
   BR$_{W\to {\rm had}}$ &/ & / & -0.0200  & / & /& /\\     \hline
$\sin^2 \theta^{\text{lep}}_{\text{eff}}$ &0.0483&-0.0881&0.115&-0.176&-0.0612& -0.0698\\  \hline
   $\mu_{ggh}^{\gamma\gamma}/\mu_{\rm SM}$ & 5.8 & / & /  & / & /& /\\     \hline\hline
    \end{tabular}}
  \caption{The coefficients for calculating the 6D-operator contributions to all the observables~\cite{Chiu:2017yrx}. We have also neglected the effects of renormalization group running. }
  \label{tab:coeff}
\end{table}

Notably, the main EWPOs are sensitive to four linear combinations of the six Wilson coefficients only: $\xi_0=-1.1c_{WB}+2c_T-4c^{(3)l}_{L} + 4c^{(3)l}_{LL}$, $\xi_\pm = c_{L}^{(3)l} \pm c^l_L$ and $c_R^e$. Explicitly, $\sigma_{\rm had}$ is sensitive to all of them; $A_b$ and $R_b$ depend on $\xi_0$ only; $A_{\text{FB}}^{b,\ell}$, $A_\ell$, and $\sin^2 \theta_{\text{eff}}^{\text{lep}}$ have the same dependence on $\xi_0$, $\xi_+$ and $c_R^e$; and $R_{\ell}$ depends on $\xi_0$, $\xi_+$ and $c_R^e$ uniquely. This necessarily leaves two (approximately) runaway directions. These directions could be partly lifted by $\Gamma_Z$ and the three $W$ observables, which have different dependences on the variables beyond $\xi_{0,\pm}$ and $c_R^e$. In addition, we will include the relative Higgs diphoton signal strength in the gluon fusion channel, namely $\mu_{ggh}^{\gamma\gamma}/\mu_{\rm SM}$, to enhance the constraint on $\mathcal{O}_{WB}$.  Only the ATLAS measurements~\cite{Rossi:2022alp} will be considered here. The CMS results are similar~\cite{CMS:2020omd} and combining them will not qualitatively change the conclusion. With this set of observables, all runaway directions can be lifted, which we will discuss more in next section.

In this paper, we plan to study the potential to further test the relevant SMEFT at future circular $e^-e^+$ colliders such as FCC-ee~\cite{FCC:2018byv} and CEPC~\cite{CEPCStudyGroup:2018ghi} also. These machines, with extraordinary integrated luminosities and state-of-the-art detector technologies, would provide unprecedented precisions for measuring $m_W$ and other EWPOs. For concreteness, we consider the ones at FCC-ee~\cite{FCC:2018byv} and list them in Table~\ref{tab:data-lep} also. These projected precisions receive contributions from both the systematic and statistical uncertainties. The projected $m_W$ precision goes below $0.5$~MeV, improved by more than one order of magnitude compared to the CDF measurement, thanks to the dedicated beam energy scanning around the $W^+W^-$ threshold. The other EWPTs would also be improved substantially at future $e^-e^+$ colliders. Notably, we will use BR($h \to \gamma\gamma$)/BR($h \to \gamma\gamma$)$_{\rm SM}$ to replace $\mu_{ggh}^{\gamma \gamma}/\mu_{\rm SM}$ here as they are mutually equivalent given the six SMEFT operators in Eq.~\eqref{eq:opes} while the latter is not directly measurable at $e^-e^+$ colliders. Such a choice also allows us to combine the HL-LHC and future $e^-e^+$ collider data for a further precision improvement~\cite{Fan:2014vta}.

At a future $e^-e^+$ collider, the theoretical uncertainties for the considered EWPOs become more relevant. Although not playing a significant role at the current stage, these theoretical uncertainties may become comparable to the experimental ones as the latter are expected to be improved more by the time of running a future $e^-e^+$ collider. For most of the relevant EWPOs, the major factors determining their theoretical uncertainties by that time, namely the expected input parameter precisions and the intrinsic uncertainties in relation to higher-order corrections, have been discussed in~\cite{dEnterria:2016rbf,Freitas:2019bre}. The exceptions are $\sigma_{\rm had}^0$, $\Gamma_W$ and BR$_{W\to{\rm had}}$. For these observables,  no future projections for their intrinsic uncertainties from the higher-order corrections are available. So we adopt a conservative assumption that their future values will be reduced by half compared to current ones. The projected theoretical uncertainties are summarized in Table~\ref{tab:data-lep}.

\section{Analysis of SMEFT}
\label{sec:results}

\subsection{General Analysis}

\begin{figure}[h]
\centering
\includegraphics[height=7.1 cm]{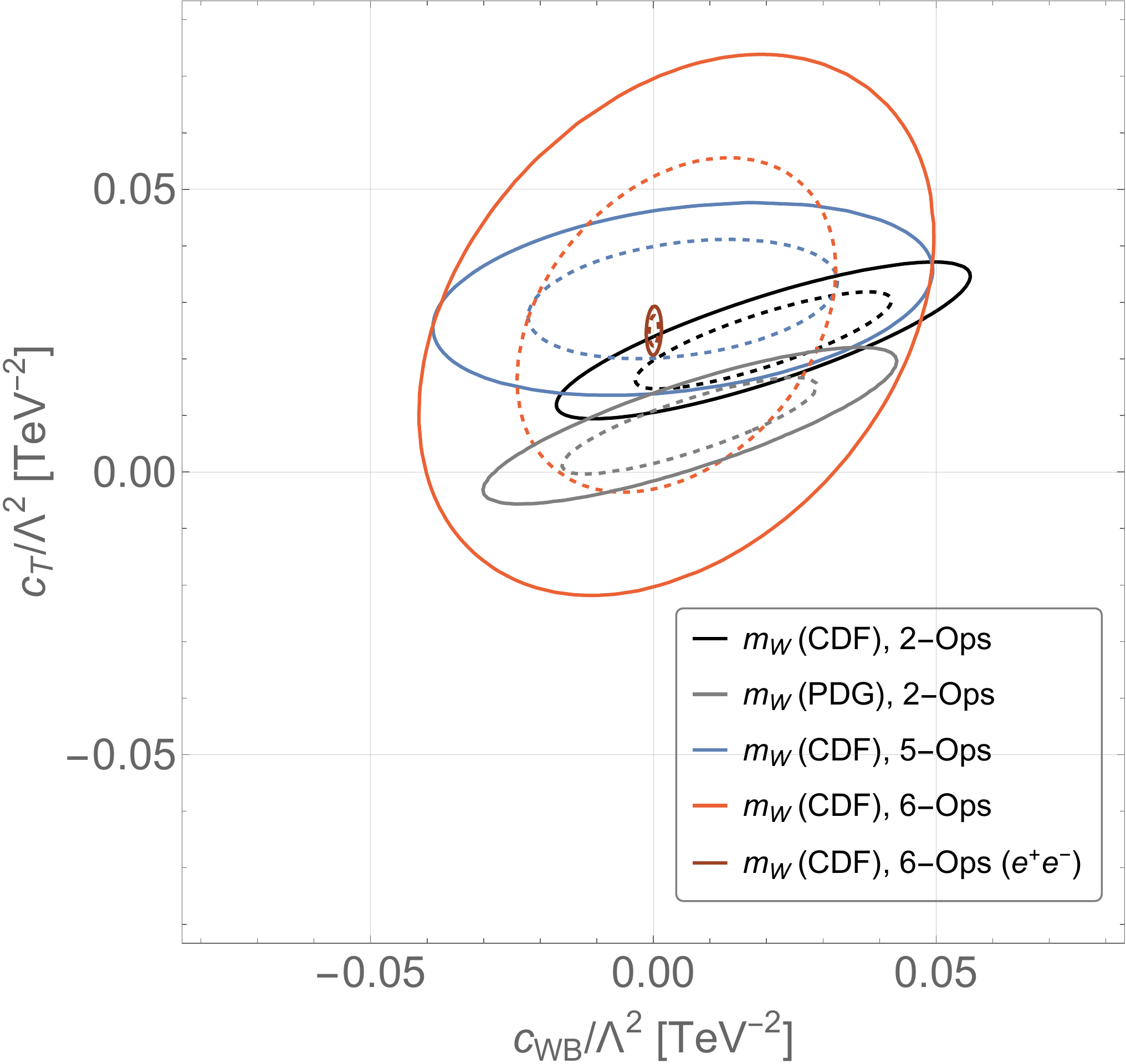}
\includegraphics[height=6.8 cm]{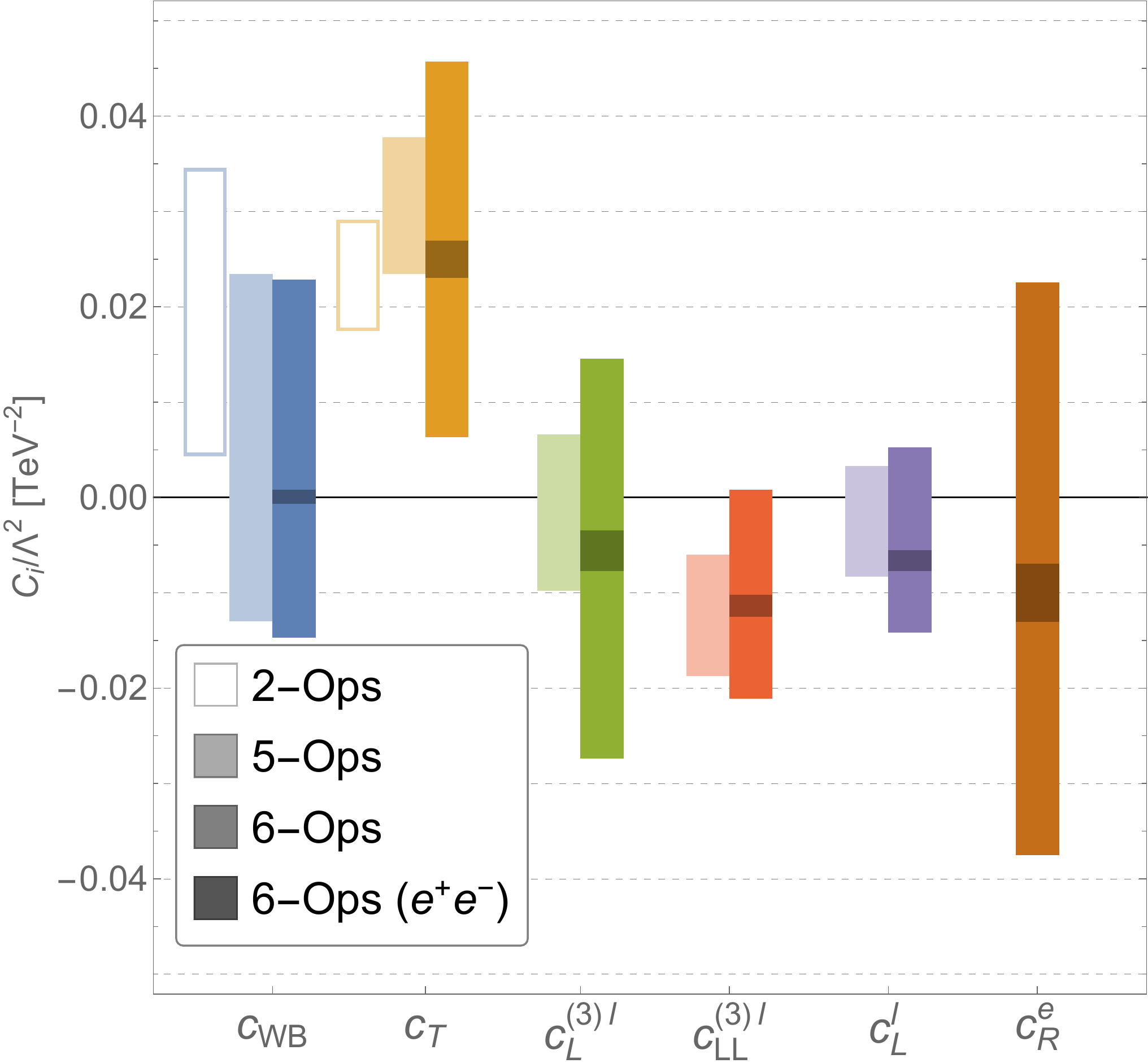}
\caption{{\textbf{Upper:}} Posterior distributions of the Wilson coefficients $c_{WB}$ and $c_{T}$ in five fitting scenarios. The dashed and solid contours are defined at 68\% and 95\% C.L., respectively. The contours have been marginalized in the five- and six-operator fitting scenarios. For the six-operator case, we have also plotted the future $e^-e^+$ collider projections as the dark orange contours. {\textbf{Bottom:}} Marginalized constraints for individual Wilson coefficients at 68\% C.L. in the fitting scenarios of the two-, five-, six-operator with current precision, and of six-operator at a future $e^-e^+$ collider.}   \label{fig:two}
\end{figure}

\begin{figure*}[th]
\centering
\includegraphics[width=12 cm]{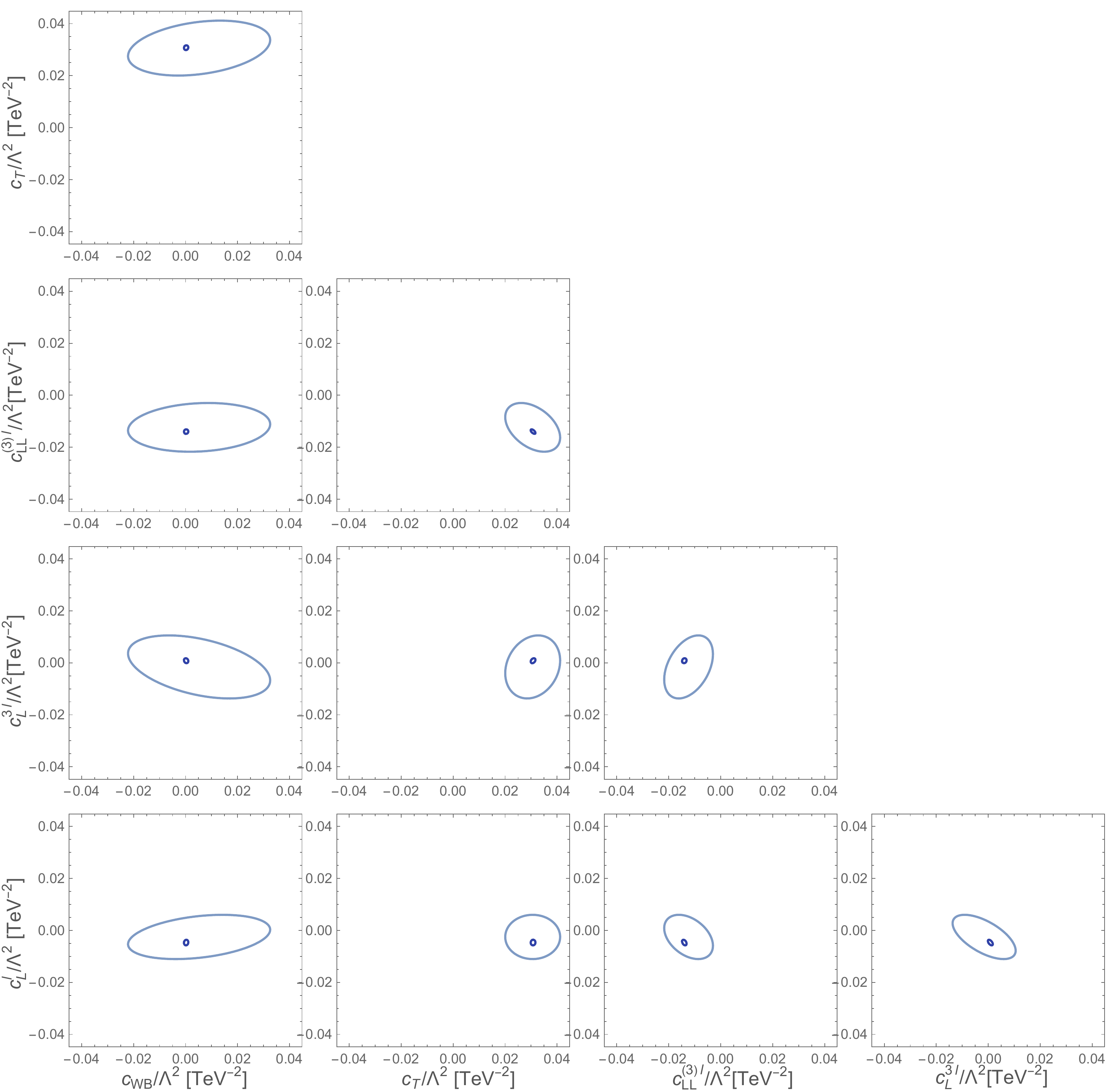}
\caption{2D posterior distributions of Wilson coefficients (\textit{Case (2)}), obtained from the marginalized $\chi^2$ analysis. The contours based on the current precisions and the precisions from a future $e^-e^+$ collider are drawn in cyan and dark cyan, respectively.}
\label{fig:tri5D}
\end{figure*}

\begin{figure*}[th]
\centering
\includegraphics[width=12 cm]{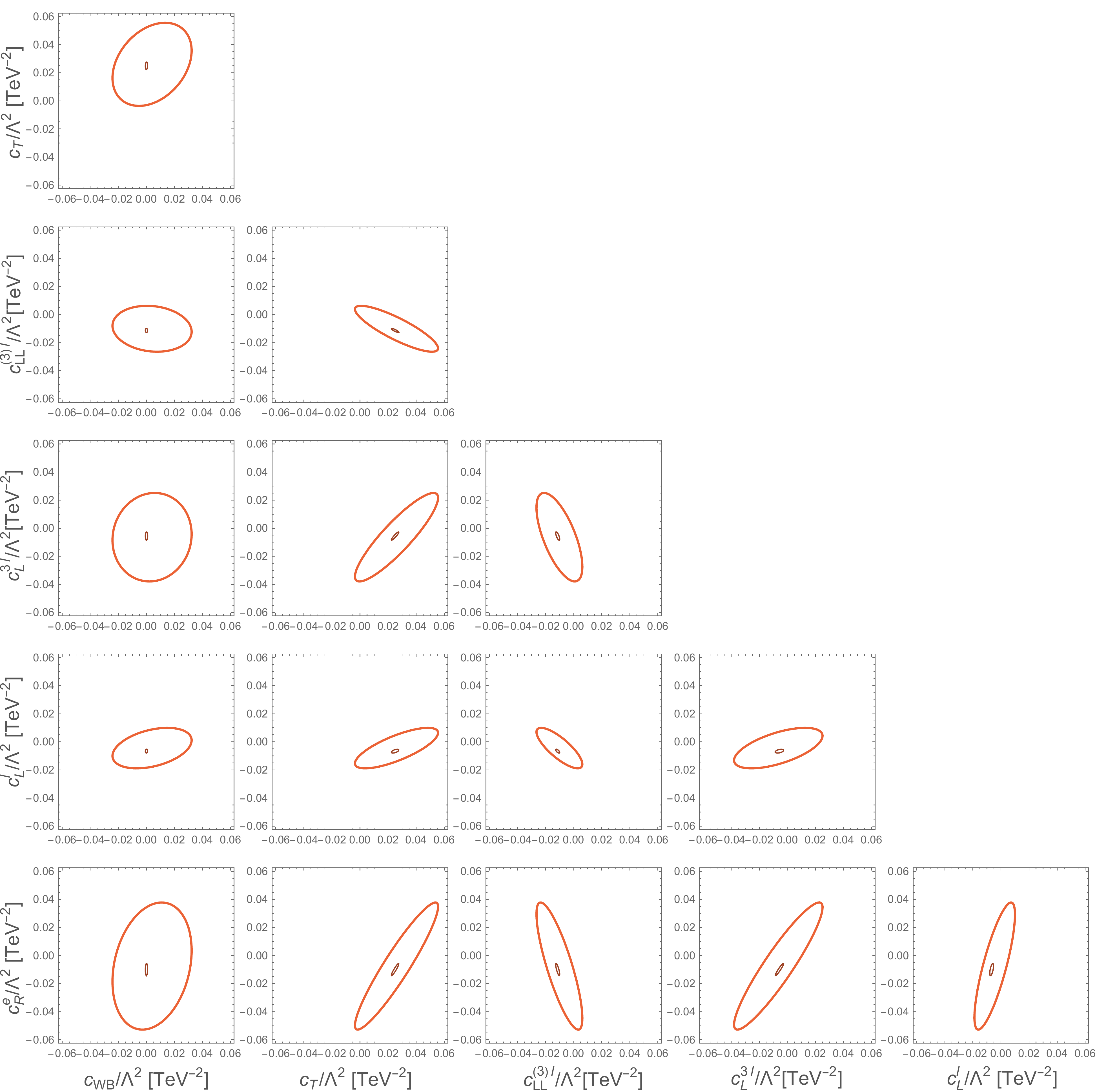}
\caption{2D posterior distributions of Wilson coefficients (\textit{Case (3)}), obtained from the marginalized $\chi^2$ analysis. The contours based on the current precisions and the precisions from a future $e^-e^+$ collider are drawn in orange and dark orange, respectively.}
\label{fig:tri6D}
\end{figure*}

Below we will analyze the impacts of the CDF $m_W$ value and other EWPTs on the set of 6D operators listed in Eq.~(\ref{eq:opes}). We will fit the data in three cases, where (1) only $\mathcal {O}_{WB}$ and $\mathcal {O}_{T}$, (2) all of the six operators except $\mathcal {O}_{R}^e$, and (3) all of the six operators are turned on. Predictions of all the observables at the best-fit points are summarized in Tab.~\ref{tab:best-fit}. The values of the Wilson coefficients favored by the data are shown in Fig.~\ref{fig:two}. For completeness, we also present the 2D posterior distributions of these Wilson coefficients in Fig.~\ref{fig:tri5D} (case (2)) and Fig.~\ref{fig:tri6D} (case (3)).

\begin{table}[ht]
  \centering
\scalebox{0.9}{
    \begin{tabular}{c|c|c|c}
    \hline
    \hline
   Observables  & Case (1) & Case (2) & Case (3) \\
    \hline
    \hline
    $m_W$(GeV)   & 80.4182  & 80.4335   & 80.4335\\
    \hline
    $A_b$    & 0.934895  & 0.93481 &  0.934944\\
    \hline
   $A_\ell$ ($P_{\tau}$)    &  0.14889  & 0.14744  & 0.14736\\
    \hline
    $A_\ell$ (SLD)    &  0.14889   & 0.14744 & 0.14736\\
    \hline
    $R_b$    &  0.21587 & 0.21588 & 0.21587 \\
    \hline
    $R_\ell$   & 20.7510 &  20.7592 &  20.7634 \\
    \hline
     $A_{\rm FB}^b$ & 0.10448 & 0.10340   & 0.10335 \\
    \hline
         $A_{\rm FB}^\ell$ &  0.01657  & 0.01629 & 0.01627\\
    \hline
    $\Gamma_Z$(GeV) &  2.49818  & 2.49515 & 2.49537\\
      \hline
    $\sigma_{\rm had}^0$(nb) & 41.4915   &  41.4729  &  41.4771 \\
      \hline 
    $\Gamma_W$(GeV) & 2.09262  & 2.09109 & 2.09261\\
      \hline
  BR$_{W\to {\rm had}}$ & 0.6748   &  0.6748 &  0.6749 \\ 
  \hline
    $\sin^2 \theta_{\text{eff}}^{\text{lep}} (10^{-5})$  &  23127.7    &  23146.6 & 23147.7 \\ 
    \hline
    $\mu_{ggh}^{\gamma\gamma}/\mu_{\rm SM}$ & 1.11 & 1.03 & 1.02 \\
    \hline
    \hline
    $\chi^2/{\rm D.O.F}$   & 1.38 & 1.20 & 1.34\\
    \hline
    \hline
    \end{tabular}}%
  \caption{Values of the EWPOs and Higgs observable at the best-fit points and corresponding $\chi^2/{\rm D.O.F}$.} 
  \label{tab:best-fit}%
\end{table}%

{\it Case (1): only $\mathcal {O}_{WB}$ and $\mathcal {O}_{T}$ are turned on.} The results are presented as black (with the CDF $m_W$ value~\cite{doi:10.1126/science.abk1781}) and gray (with the global average value~\cite{ParticleDataGroup:2020ssz}) contours in Fig.~\ref{fig:two}. This is equivalent to the standard EW fit with new physics encoded in the oblique parameters $S$ and $T$~\cite{Kennedy:1988sn,Holdom:1990tc,Peskin:1990zt, Peskin:1991sw,Golden:1990ig}. Here $c_{WB}$ and $c_{T}$ are related to the $S$ and $T$ parameters as
\begin{eqnarray}
S &=& 16\pi \frac{c_{WB} \, v^2}{\Lambda^2} \approx 3 c_{WB}\left(\frac{{\rm TeV}}{\Lambda}\right)^2 , \nonumber \\
T &=& \frac{c_T}{\alpha} \frac{v^2}{\Lambda^2} \approx 7.7 c_{T}  \left(\frac{{\rm TeV}}{\Lambda}\right)^2 \ ,
\end{eqnarray}
where $v = 246$ GeV. 
It is clear that with the global average of $m_W$, the contours are approximately centered at $\{0,0\}$ as expected, while the contours with the CDF $m_W$ value take a right-upward offset from the $\{0,0\}$ point such that a positive $c_T$ is strongly favored. The latter yields $c_T ({\rm TeV}/\Lambda)^2 \gtrsim 0.015$ at 68\% C.L. Combining Tab.~\ref{tab:coeff} and Eq.~(\ref{eq:ge}), we have
\begin{equation}
\begin{split}
\dfrac{\Delta m_W}{m_W} =&  - 0.0111 \dfrac{c_{WB}}{\Lambda^2}  +0.0433 \dfrac{c_{T}}{\Lambda^2}  \\
&-
 0.0264 \dfrac{c^{(3)l}_L}{\Lambda^2} +0.0264 \dfrac{c^{(3)l}_{LL}}{\Lambda^2}
\end{split} \ .
\end{equation}
A shift in $m_W$ required for explaining the CDF measurement is thus essentially driven by the operator $\mathcal {O}_{T}$. This conclusion can be extended to other cases also. 

{\it Case (2):  all of the six operators except $\mathcal {O}_{R}^e$ are turned on.} The results are presented as blue contours in the left panel of Fig.~\ref{fig:two} and the light-shaded bars in its right panel. In the left panel, we have marginalized the other three coefficients to obtain the contours in the $(c_{WB}, c_T)$ plane. Compared to the 2D fit in the previous case, the contours expand as expected, corresponding to a larger allowed parameter space. Yet an even larger positive $c_T$ is needed to explain the CDF $m_W$ value. This point can also be seen from the right panel. We also provide all the 2D posterior distributions of the five Wilson coefficients in Fig.~\ref{fig:tri5D}.

{\it Case (3): all of the six operators are turned on.}  The results are presented as orange contours in the left panel of Fig.~\ref{fig:two} and the dark shaded bars in its right panel. The marginalized contours in the 2D space and the 1D bars continue to expand in this case. However, at 68\% C.L., $c_T ({\rm TeV}/\Lambda)^2 \gtrsim 0.01$ is favored after a full marginalization, as it happens to the other cases. The 2D posterior distributions of the six Wilson coefficients are presented in Fig.~\ref{fig:tri6D}. We could see correlations between several Wilson coefficients, such as $c_R^e$, $c_T$ and $c_L^{3l}$.

\subsection{Discussions}

In general, $c_T ({\rm TeV}/\Lambda)^2 \gtrsim 0.01$ is needed to accommodate the large $m_W$ value and its high precision reported by CDF. These results may not be easily explained by the BSM physics, even if we ignore the tension between the CDF result and other collider measurements. 
Firstly, the relevant cutoff scale tends to be low. If this operator is generated at one-loop level in a BSM scenario, the new mass scale is $\sim {\cal O}(100)$ GeV, assuming order-one couplings. If this operator is generated at tree level, the new mass scale could be raised to a few TeV. In either case, the new physics scale could be well within the reach of current and near-future LHC searches.
Secondly, $c_T$ has to be positive to increase $m_W$. This may not always be the case in BSM scenarios that could change $m_W$. As an example, depending on the symmetry breaking pattern, some little Higgs models shift $m_W$ in the opposite direction~\cite{Han:2005dz, Giudice:2007fh, Chiu:2017yrx}. Combining these two considerations, it is easier to seek a tree-level explanation for the large $m_W$. One example is augmenting the SM with a single electroweak triplet scalar field with zero hypercharge. In this case, $m_W$ receives a positive contribution at tree level~\cite{Khandker:2012zu, Ellis:2020unq}, while the contribution to $c_{WB}$ arises at one-loop order. More work needs to be done to exhaust all possible models and check other constraints on them, which we will not explore further in this article. Note that our analysis does not apply to low-scale BSM scenarios where the SMEFT fails.

Separately, the physical origin of this $m_W$ anomaly can be further tested indirectly with the observables which are sensitive on these operators. Such observables typically exist for the EW processes at colliders, such as di-boson productions and vector-boson fusion scatterings. Moreover, one may consider the partial decay of Higgs boson to diphoton. As shown in the data fits (especially in case (1)), $c_{WB}$ is allowed to vary at $\mathcal O(0.01)$. Because of its high-sensitivity to $c_{WB}$ (since such decays arise from one-loop level in the SM; see Tab.~\ref{tab:coeff} also), such $c_{WB}$ values would lead to a deviation in $\Gamma_{h\to \gamma\gamma}$ at $\sim 10\%$ level. This is already comparable to the current precision at the LHC. At HL-LHC, this precision is expected to reach a level $\sim 4\%$ in the gluon fusion channel~\cite{Cepeda:2019klc}. Such measurements certainly will probe the BSM physics along the $c_{WB}$ direction (constraining $|c_{WB}| ({\rm TeV}/\Lambda)^2 \lesssim 0.007$) and hence narrow the allowed space for explaining the $m_W$ anomaly.

Compared to the LHC and HL-LHC, the future $e^-e^+$ colliders can provide a more decisive test of the relevant SMEFT. To demonstrate this point, we present the marginalized 1D constraints and 2D posterior distributions at FCC-ee for the Wilson coefficients in Case (3) in Fig.~\ref{fig:two} (right panel) and Fig.~\ref{fig:tri6D} (also see Fig.~\ref{fig:two} (left panel)), respectively. We also present the 2D posterior distributions at FCC-ee for the Wilson coefficients in Case (2) in Fig.~\ref{fig:tri5D}. Here the future experimental central values for the relevant observables have been assumed to have no shift from their current ones (see Tab.~\ref{tab:data-lep}). The FCC-ee constraints are much stronger than the current ones, thanks to the improvements of both the experimental and theoretical uncertainties. In particular, from Fig.~\ref{fig:two} (right panel), one can see that $|c_{WB}|$ is tightly constrained to be $\lesssim 10^{-3}$, while $c_T$, as the potential main cause of the $M_W$ anomaly, can be tested with more than $10\sigma$ away from its null limit. At last, we expect comparable sensitivities to be achieved at CEPC.

\section{Summary}
\label{sec:conclusions}

Our paper explores the possible new physics origin of the recent CDF $m_W$ measurement, focusing on alleviating the tension between the CDF result and EWPTs. We carry out a model-independent analysis using a subset of 6D operators which EWPOs are most sensitive to, in the SMEFT framework. We implement three different fits and show that to accommodate the CDF result, new physics has to generate $\mathcal{O}_{T}=\frac{1}{2}(H^{\dagger}\overset{\text{$\leftrightarrow$}}{D}_{\mu}H)^2$ with a coefficient $c_T ({\rm TeV}/\Lambda)^2 \gtrsim 0.01$. This suggests a new energy scale of multiple TeV for tree-level effects and sub TeV for loop-level effects, which could be searched for either through direct searches or other indirect probes. 

While the CDF result clearly needs to be cross-checked with the upcoming LHC measurements, it certainly urges the particle physics community to prepare better for understanding the potential physical cause of the anomalies from indirect precision measurements. This could be crucial for the planning of future collider projects. As we have shown, a future $e^-e^+$ collider covering the $Z$-pole, $WW$ threshold, Higgs factory, and $t\bar{t}$ threshold modes could be highly relevant or even decisive in this regard.\\

\begin{acknowledgments}

We thank Ayres Freitas, Ashutosh Kotwal, Kirtimaan Mohan, Simone Pagan Griso, and Keping Xie for useful discussions. J. Fan and L. Li are supported by the DOE grant DE-SC-0010010. T. Liu is supported partly by the Area of Excellence (AoE) under the Grant No.~AoE/P-404/18-3, and partly by the General Research Fund (GRF) under Grant No.~16305219. Both the AoE and GRF grants are issued by the Research Grants Council of Hong Kong S.A.R. K. Lyu is partially supported by the DOE grant DE-SC0022345.

\end{acknowledgments}

\bibliographystyle{apsrev}
\bibliography{main.bib}

\end{document}